\documentclass[aip,
nofootinbib,
prl,%
reprint,% single-spaced, dense output
twocolumn,
longbibliography,
author-numerical
]{revtex4-1}

\usepackage{amsfonts,amssymb,amsmath}
\usepackage[usenames]{color}
\usepackage{graphicx}
\usepackage{natbib}
\usepackage{changes}
\usepackage{braket}

\newcommand{\pdr}[2]{\dfrac{\partial   {#1}}{\partial {#2}}}
\newcommand{\pddr}[2]{\dfrac{\partial^2 {#1}}{\partial {#2}^2}}

\newcommand{\tx}{\tilde{x}}
\newcommand{\ty}{\tilde{y}}
\newcommand{\tw}{\tilde{w}}
\newcommand{\tu}{\tilde{u}}
\newcommand{\tv}{\tilde{v}}

\newcommand{\tp}{\tilde{p}}

\newcommand{\cref}{c_h^0}

\newcommand{\Reyn}{{\rm Re}}

\newcommand{\tit}{\tilde{t}}
\newcommand{\tom}{\tilde{\omega}}
\newcommand{\ri}{{\rm i}}

\renewcommand{\Re}[1]{\operatorname{Re}\left(#1\right)}

\begin{document}

% \linenumbers

\sf
\title{Oscillations of laminar flow velocity in a channel induced
    by harmonic perturbation of mass injection velocity through the permeable wall}
\author{Andrei Kulikovsky}
% \thanks{ECS member}
\email{A.Kulikovsky@fz-juelich.de}
\affiliation{Forschungszentrum J\"ulich GmbH           \\
     Theory and Computation of Energy Materials (IEK--13)   \\
     Institute of Energy and Climate Research,              \\
     D--52425 J\"ulich, Germany
 }
\altaffiliation[Also at: ]{Lomonosov Moscow State University,
     Research Computing Center, 119991 Moscow, Russia}
\date{\today}

\date{\today}

\begin{abstract}
    Transient 2D Navier--Stokes equations for the laminar flow of incompressible fluid in a channel
    with permeable wall are reduced to a single transient one--dimensional
    equation for the transversal profile of longitudinal
    flow velocity. Small--amplitude harmonic perturbation of injection
    velocity induces oscillations of longitudinal velocity with the peak at the walls.
    The peak amplitude increases with the distance along the channel.
    The effect of increasing amplitude is explained by formation of linearly growing along
    the channel perturbation
    of pressure gradient. With the frequency growth, the peaks move toward to the walls.
\end{abstract}


\maketitle

\section{Introduction}

Laminar flow of incompressible fluid in channels with permeable wall(s) is of interest
for ultrafiltration applications~\citep{Chao_18} and fuel cells~\citep{EK_book_14}.
In 1953, \citet{Berman_53} reduced 2D problem for the flow between parallel
permeable walls with constant velocity of injection to
a single ODE for the transversal shape of longitudinal velocity and provided
an elegant asymptotic solution. Later, Berman's approach has been used
to solve the problem of a flow in pipes and ducts
with permeable walls and variable along the pipe/duct velocity
of suction/injection~(\citet{Terrill_69,Terrill_83,Kosinski_70,Granger_89,Karode_01}).
So far, however, the transient effects due to time--dependent
injection velocity have not been considered.

In 1929, \citet{Richardson_1929} reported measurements
of oscillating flow in a pipe with impermeable wall
induced by harmonic longitudinal pressure gradient.
They demonstrated formation of a peak of velocity oscillations
amplitude close to the pipe wall. A year later, \citet{Sexl_1930}
developed a model for oscillating flow in a circular pipe and derived a
simple solution for the radial shape of velocity amplitude.
He showed that the oscillations amplitude is distributed along
the radius according to the Bessel function with a peak (shoulder)
positioned at the distance
\begin{equation}
   l_* = \sqrt{\dfrac{\nu}{\omega}}
   \label{eq:last}
\end{equation}
from the wall.
Here
$\nu$ is the air kinematic viscosity and
$\omega$ is the angular frequency of perturbation.
\citet{Harris_69} provided accurate measurements
confirming the result of Sexl. In all these works the velocity oscillations
were induced by harmonic variation of longitudinal pressure gradient.

Below, unsteady laminar flow of incompressible fluid
between parallel walls with a variable and time--dependent rate of mass
injection through one of the walls is considered.
Following the Berman's approach, the system of transient Navier--Stokes equations
is reduced to a single equation for the transversal shape
of longitudinal flow velocity. The equation is used to study the
flow response to a small--amplitude harmonic perturbation of the injection
velocity. It is shown that such a perturbation induces
perturbations of the longitudinal flow velocity.
At high frequencies, close to the walls a peak of oscillations amplitude forms
which increases dramatically with the distance along the channel.
The effect is explained by linearly growing along the channel perturbation
amplitude of pressure gradient. The peaks are located at the distance
$l_*$ from the nearest wall.

\section{Model}

\subsection{Basic equations}

\begin{figure}
    \begin{center}
        \includegraphics[scale=0.6]{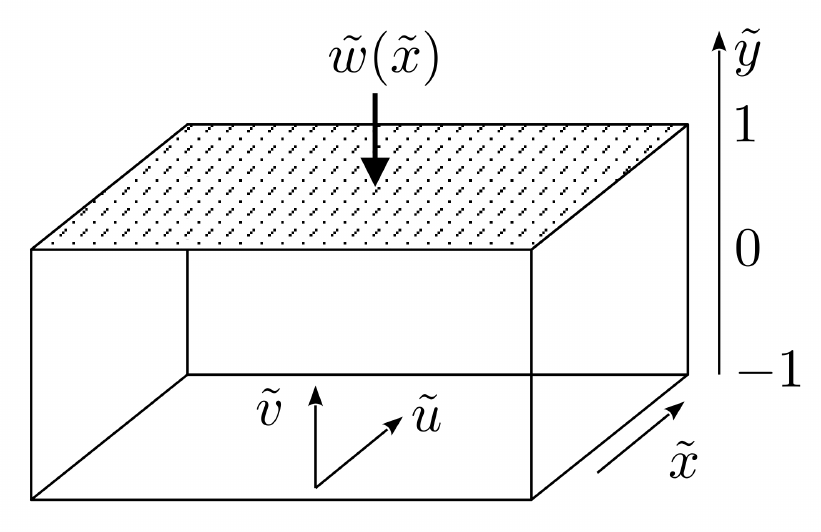}
        \caption{Schematic of the channel.
            Hatched area indicates the permeable wall through
            which the mass enters the channel.
        }
        \label{fig:slit}
    \end{center}
\end{figure}

Consider laminar flow of incompressible fluid between parallel walls
separated by the distance $2h$ and let the upper wall be permeable
to mass injection (Figure~\ref{fig:slit}).
Navier--Stokes equations for longitudinal $u$ and transversal $v$
flow velocity components are
\begin{equation}
   \pdr{u}{t} +  u\pdr{u}{x} + v\pdr{u}{y}
    = -\dfrac{1}{\rho}\pdr{p}{x} + \nu\left(\pddr{u}{x} + \pddr{u}{y}\right)
    \label{eq:NSux}
\end{equation}
\begin{equation}
   \pdr{v}{t} + u\pdr{v}{x} + v\pdr{v}{y}
    = -\dfrac{1}{\rho}\pdr{p}{y} + \nu\left(\pddr{v}{x} + \pddr{v}{y}\right)
    \label{eq:NSvy}
\end{equation}
Here $p$ is the pressure and $\rho$ is the flow density.
As $\rho$ is constant, the continuity equation is
\begin{equation}
    \pdr{u}{x} + \pdr{v}{y} = 0
    \label{eq:cont}
\end{equation}
Introducing dimensionless variables
\begin{multline}
    \tit = \dfrac{t u_0}{h}, \quad \tx = \dfrac{x}{h}, \quad \ty = \dfrac{y}{h}, \quad
    \tu = \dfrac{u}{u_0}, \quad \tv = \dfrac{v}{u_0}, \\
    \tp = \dfrac{p}{\rho u_0^2}, \quad \tom = \dfrac{\omega h}{u_0},
    \label{eq:dless}
\end{multline}
Eqs.\eqref{eq:NSux} -- \eqref{eq:cont} transform to
\begin{equation}
    \pdr{\tu}{\tit} + \tu\pdr{\tu}{\tx} + \tv\pdr{\tu}{\ty}
    = - \pdr{\tp}{\tx} + \dfrac{1}{\Reyn}\left(\pddr{\tu}{\tx} + \pddr{\tu}{\ty}\right)
    \label{eq:tNSux2}
\end{equation}
\begin{equation}
    \pdr{\tv}{\tit} + \tu\pdr{\tv}{\tx} + \tv\pdr{\tv}{\ty}
    = -\pdr{\tp}{\ty} + \dfrac{1}{\Reyn}\left(\pddr{\tv}{\tx} + \pddr{\tv}{\ty}\right)
    \label{eq:tNSvy2}
\end{equation}
\begin{equation}
    \pdr{\tu}{\tx} + \pdr{\tv}{\ty} = 0,
    \label{eq:tcont}
\end{equation}
where
\begin{equation}
    \Reyn = \dfrac{u_0 h}{\nu}
    \label{eq:Rey}
\end{equation}
is the inlet Reynolds number,
$u_0$ is the mean over the $y$--axis $x$--component of inlet flow velocity,
and $\omega$ is the angular frequency of perturbations (see below).

Following the idea of \citet{Berman_53} we introduce a stream function
\begin{equation}
    \psi = \left(1 + \int_0^{\tx} \tw\left(\tit, \xi\right)\, d\xi \right)
        f\left(\tit, \ty\right)
    \label{eq:psi}
\end{equation}
where $\tw$ is the dimensionless velocity of injection. Setting
\begin{equation}
    \tu = \pdr{\psi}{\ty}, \quad \tv = - \pdr{\psi}{\tx},
    \label{eq:CR}
\end{equation}
Eq.\eqref{eq:tcont} is satisfied. For $\tu$ and $\tv$ we thus have
\begin{equation}
    \tu = (1 + R) f', \quad \tv = - \tw f
    \label{eq:tutv}
\end{equation}
where
\begin{equation}
    R \equiv \int_0^{\tx} \tw\left(\tit, \xi\right)\, d\xi,
    \label{eq:R}
\end{equation}
the prime sign indicates partial derivative over $\ty$ or $\tx$,
depending upon the variable
\begin{equation}
   f' \equiv \pdr{f}{\ty}, \quad \tw' \equiv \pdr{\tw}{\tx},
   \label{eq:prime}
\end{equation}
and similar for higher derivatives.
Substituting Eqs.\eqref{eq:tutv} into Eqs.\eqref{eq:tNSux2}, \eqref{eq:tNSvy2} we come to
\begin{multline}
    \pdr{(1 + R) f'}{\tit} + (1 + R) \tw \left(f'f' -  f f''\right) \\
    = - \pdr{\tp}{\tx} + \dfrac{1}{\Reyn}\left(\tw' f' + (1 + R) f'''\right)
    \label{eq:fx1}
\end{multline}
\begin{multline}
    - \pdr{(\tw f)}{\tit} + \left( \tw^2 - (1 + R) \tw'\right) f f' \\
    =  - \pdr{\tp}{\ty} - \dfrac{1}{\Reyn}\left( \tw'' f + \tw f''\right)
    \label{eq:fy1}
\end{multline}
Differentiating Eq.\eqref{eq:fx1} over $\ty$ and Eq.\eqref{eq:fy1} over $\tx$ we get
\begin{multline}
    \pdr{(1 + R)f''}{\tit} + (1 + R) \tw \left(f'f'' - f f''' \right) \\
    = - \pdr{}{\ty}\left(\pdr{\tp}{\tx}\right)
         + \dfrac{1}{\Reyn}\bigl(\tw' f'' + (1 + R) f''''\bigr)
    \label{eq:fx2}
\end{multline}
\begin{multline}
    - \pdr{(\tw'  f)}{\tit} + \bigl( \tw\tw' - (1 + R) \tw''\bigr) f f'   \\
    =  - \pdr{}{\tx}\left(\pdr{\tp}{\ty}\right)
        - \dfrac{1}{\Reyn}\left( \tw''' f + \tw' f''\right)
    \label{eq:fy2}
\end{multline}
Subtracting Eq.\eqref{eq:fy2} from Eq.\eqref{eq:fx2} we come to
\begin{multline}
    \pdr{}{\tit}\biggl((1 + R) f'' + \tw' f\biggr)
    + (1 + R) \tw \left(f'f'' - f f''' \right) \\
    -  \bigl(\tw\tw' - (1 + R) \tw''\bigr)f f' \\
    = \dfrac{1}{\Reyn}\bigl(2\tw' f'' + (1 + R) f'''' + \tw''' f \bigr)
    \label{eq:fbase}
\end{multline}
Eq.\eqref{eq:fbase} is the general equation for the
flow in channel with a non--uniform along $\tx$
and time--dependent velocity of mass injection.
The boundary conditions for this equation follow from Eqs.\eqref{eq:tutv}:
\begin{equation}
    f(1) = 1,\quad  f'(1) = 0, \quad f(-1) = f'(-1) = 0
    \label{eq:fbc}
\end{equation}

\subsection{Small oscillations of uniform injection velocity}

It is advisable
to consider the case of uniform along $\tx$ injection velocity $\tw$.
Setting in Eq.\eqref{eq:fbase} $R = \tw(\tit)\, \tx$ and chalking out
the terms with $\tw'$, $\tw''$, and $\tw'''$, we arrive at
\begin{equation}
   \left(\dfrac{\tx}{1 + \tx\tw}\right)\pdr{\tw}{\tit} f''
    + \pdr{f''}{\tit}  + \tw \bigl(f'f'' - f f''' \bigr) = \dfrac{f''''}{\Reyn}
    \label{eq:fBerman2}
\end{equation}
Eq.\eqref{eq:fBerman2} can be integrated over $\ty$ once, leading to
\begin{equation}
    \left(\dfrac{\tx}{1 + \tx\tw}\right)\pdr{\tw}{\tit} f'
    + \pdr{f'}{\tit} + \tw \bigl(f'f' - f f'' \bigr) - \dfrac{f'''}{\Reyn} = k
     \label{eq:fBert}
\end{equation}
where $k$ is determined from solution of Eq.\eqref{eq:fBert}
with four boundary conditions, Eq.\eqref{eq:fbc}.
Eq.\eqref{eq:fBert} is the transient version of equation derived
by \citet{Berman_53}.

Substituting
\begin{equation}
    \begin{split}
       & \tw = \tw_0 + \tw_1(\tom)\exp(\ri\tom\tit), \quad |\tw_1| \ll |\tw_0| \\
       & f = f_0(\ty) + f_1(\tom, \ty)\exp(\ri\tom\tit), \quad |f_1| \ll |f_0|  \\
       & k = k_0 + k_1
    \end{split}
\end{equation}
into Eq.\eqref{eq:fBert}, neglecting terms with the perturbation products
and subtracting the static equation Eq.\eqref{eq:fBerstat},
we come to a linear problem for the complex perturbation amplitude  $f_1(\tom,\ty)$:
\begin{multline}
    \ri\tom \left(\dfrac{\tx\tw_1 f_0'}{1 + \tx\tw_0}
    + f_1'\right) + \tw_0 \bigl(2 f_0' f_1' - f_0 f_1'' - f_0'' f_1 \bigr) \\
    + \tw_1 \bigl(f_0'f_0' - f_0 f_0''\bigr) - \dfrac{f_1'''}{\Reyn} = k_1
    \label{eq:fbert1}
\end{multline}
where $f_0(\ty)$ is a solution to the static Berman's equation
\begin{equation}
    \tw_0 \left(f_0'f_0' - f_0 f_0'' \right) - \dfrac{f_0'''}{\Reyn} = k_0
    \label{eq:fBerstat}
\end{equation}
Here, the subscripts 0 and 1 mark the static variables and the small
perturbation amplitudes in the $\omega$--space, respectively.

The boundary condition for $f_1$ at $\ty=1$ follows from equation
$\tv_0 + \tv_1 = -(\tw_0 + \tw_1) (f_0 + f_1)$. At the upper wall,
$\tv_1 = - \tw_1$; neglecting the product $\tw_1 f_1$
and taking into account that $\tv_0 = - \tw_0 f_0$, $f_0(1)=1$,
we get $f_1(1) = 0$.
Thus, the boundary conditions for Eq.\eqref{eq:fbert1} are
\begin{equation}
    f_1(1) = f_1'(1) = f_1(-1) = f_1'(-1) = 0.
    \label{eq:f1bc}
\end{equation}

\section{Results and Discussion}

\begin{figure}
    \begin{center}
        \includegraphics[scale=0.45]{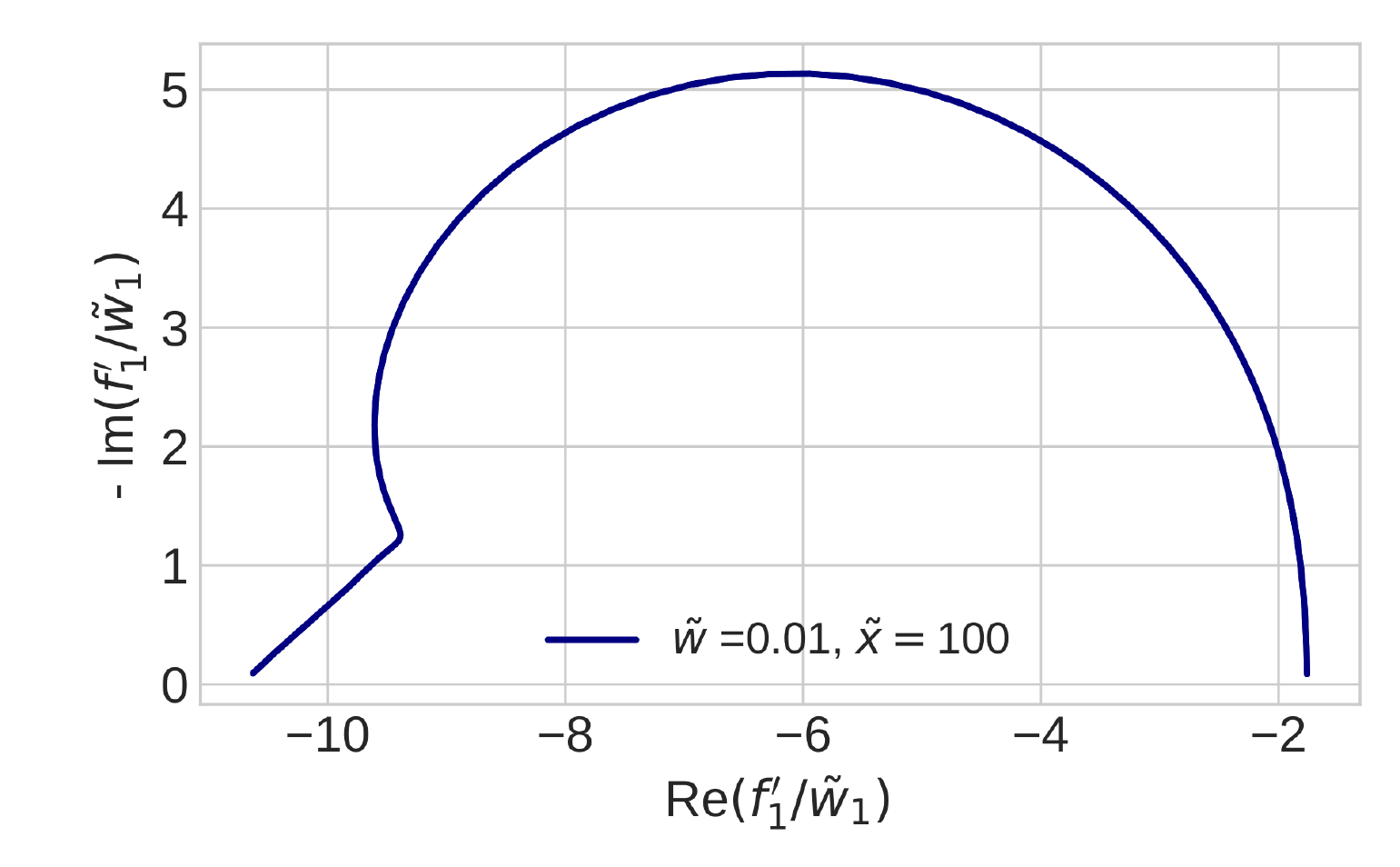}
        \caption{Nyquist spectrum of the longitudinal velocity
            oscillations amplitude $f_1' / \tw_1$
            at the mid--plane $\ty=0$ and $\tx=100$
            (one tenths of the channel length).
        }
        \label{fig:f1vw}
    \end{center}
\end{figure}

\begin{figure}
    \begin{center}
        \includegraphics[scale=0.45]{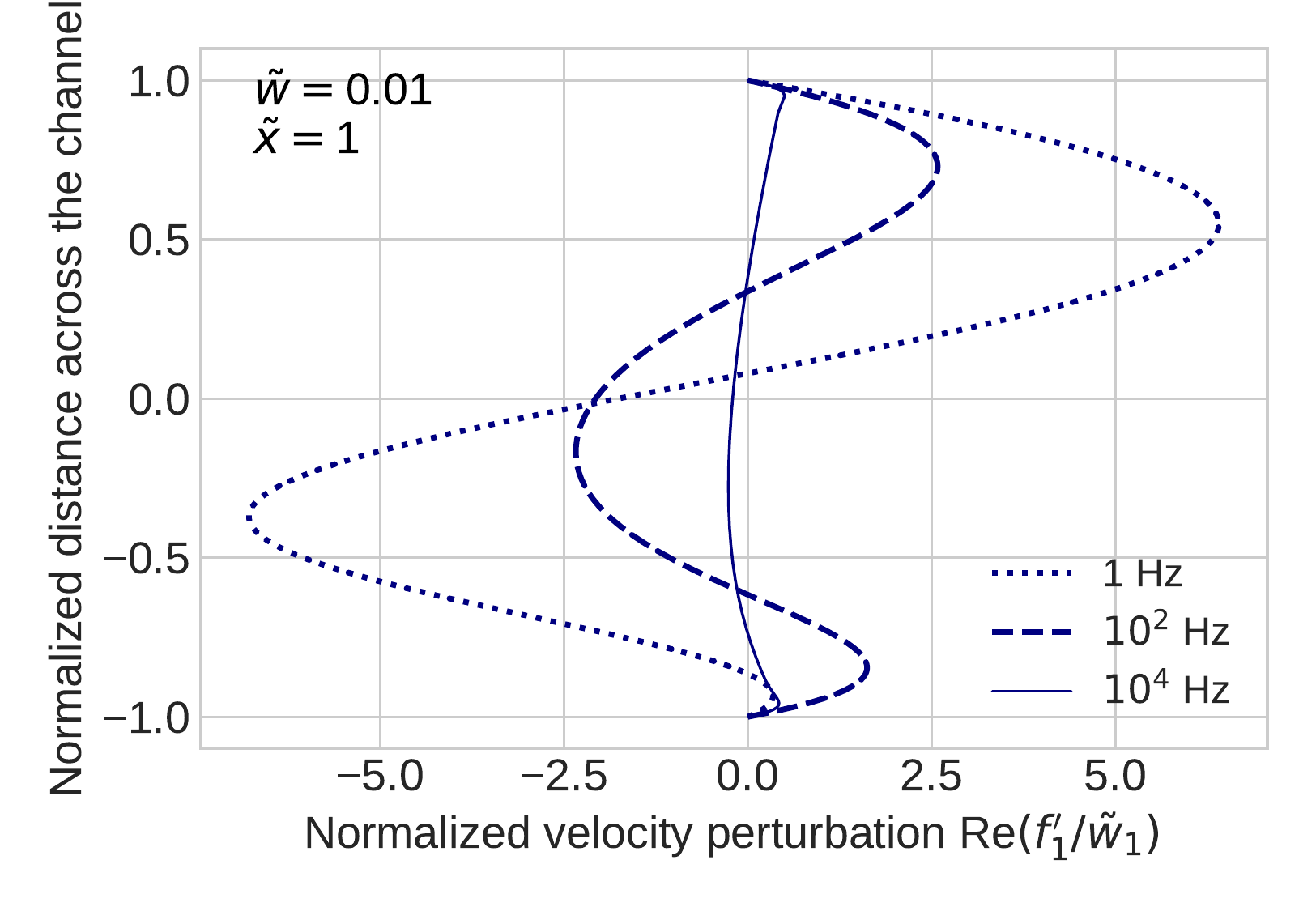}
        \includegraphics[scale=0.45]{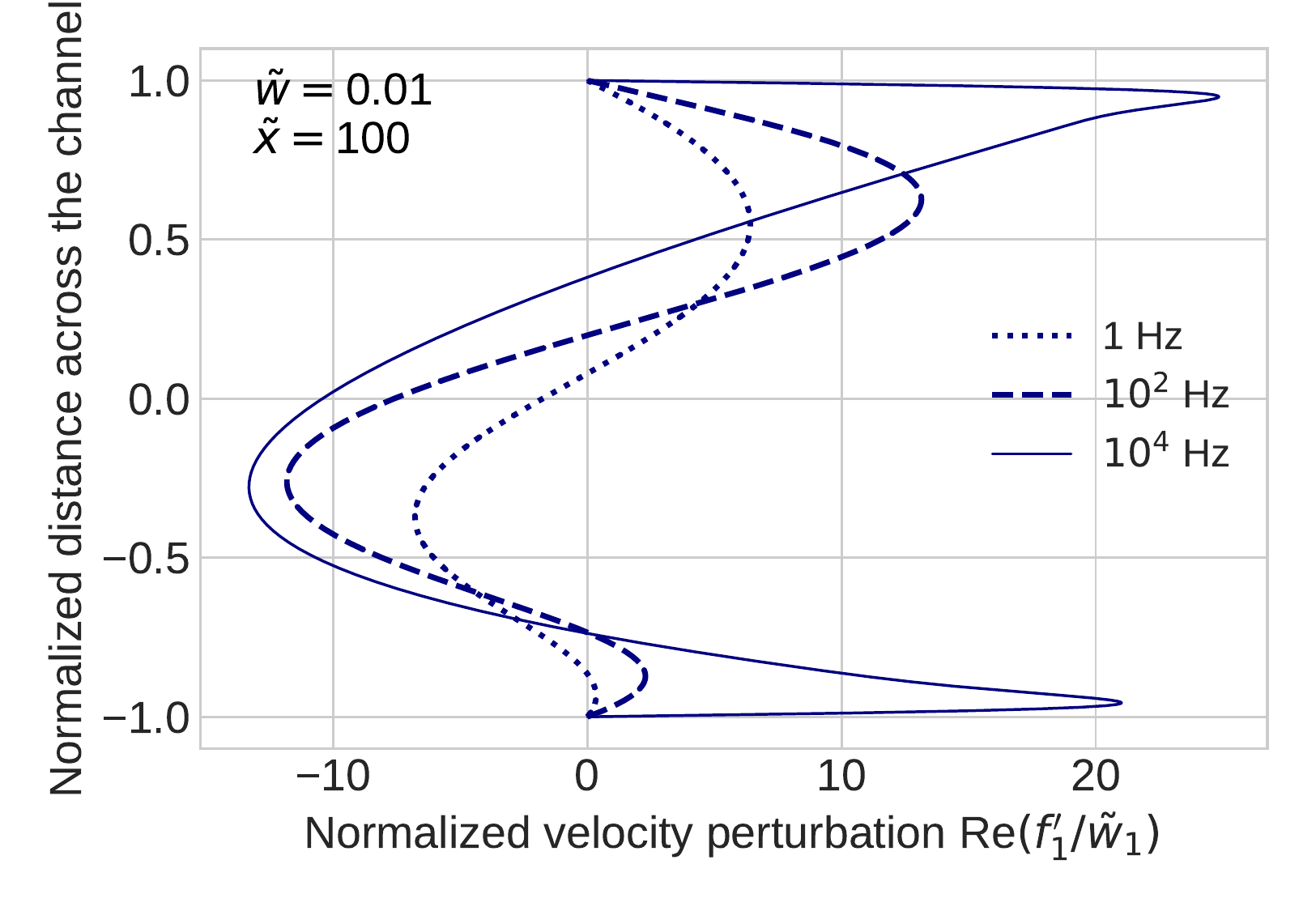}
        \caption{Transversal shape of longitudinal flow velocity
            oscillations amplitude for the indicated frequencies
            at (a) $\tx=1$ and (b) $\tx=100$
            (one tenths of the channel length).
        }
        \label{fig:utrans}
    \end{center}
\end{figure}

The spectrum of $f_1'$ at the mid--plane $\ty=0$ and $\tx=100$
for the parameters in Table~\ref{tab:parms} resembles
Warburg finite--length transport impedance~\citep{Warburg_1899} (Figure~\ref{fig:f1vw}).
Since $\tu_1 \sim f_1'$, Eq.\eqref{eq:tutv},
of particular interest is the shape of
$\Re{f_1'}$ representing the longitudinal velocity oscillations
in--phase with the applied perturbation.
Eq.\eqref{eq:fbert1} contains a term explicitly depending on $\tx$; thus, this
shape changes with $\tx$ and with the frequency. Evolution of this shape at $\tx=1$
with the growth of frequency is depicted in Figure~\ref{fig:utrans}a.
With the frequency growth,
two ``shoulders'' at the walls and a valley between them form
(the curve for $10^2$ Hz, Figure~\ref{fig:utrans}a). The amplitude of oscillation in the valley
is negative meaning that here, the oscillations decelerate the flow.
Upon further frequency growth,
the shoulders move toward the walls
(solid line in Figure~\ref{fig:utrans}a). The asymmetry of the curves
is due to mass injection at the upper wall.

\begin{table}
    \small
    \begin{center}
        \begin{tabular}{|l|c|}
            \hline
            Channel depth $2h$, m                          &  $0.1\cdot 10^{-2}$  \\
            Channel length $L$, m                          &  1.0                 \\
            Air density, $\rho$, kg~m$^{-3}$               &  1.06        \\
            Air kinematic viscosity, $\nu$, m$^2$~s$^{-1}$ & $1.886\cdot 10^{-5}$  \\
            Inlet flow velocity $u_0$, m~s$^{-1}$          & 10 \\
            Reynolds number $\Reyn$                        & 530    \\
            Injection velocity $w$, m~s$^{-1}$             & 0.1 \\
%             Perturbation amplitude of injection velocity $w_1$, m~s$^{-1}$             & 0.01 \\
            \hline
        \end{tabular}
        \caption{The flow parameters for the calculations.
        }
        \label{tab:parms}
    \end{center}
\end{table}

At $\tx=1$, the oscillations amplitude in the shoulders
is on the order of unity, i.e.,
the relative amplitude of velocity oscillations is
of the same order of magnitude as the applied amplitude of injection
velocity oscillations. However, the longitudinal velocity is two orders
of magnitude larger than the injection velocity (Table~\ref{tab:parms}),
meaning that the system works as a hydrodynamic ``amplifier''.
A close analogy is a field--effect transistor, in which a small variation of gate potential
induces large variation of the source--drain current.

\begin{figure}
    \begin{center}
        \includegraphics[scale=0.45]{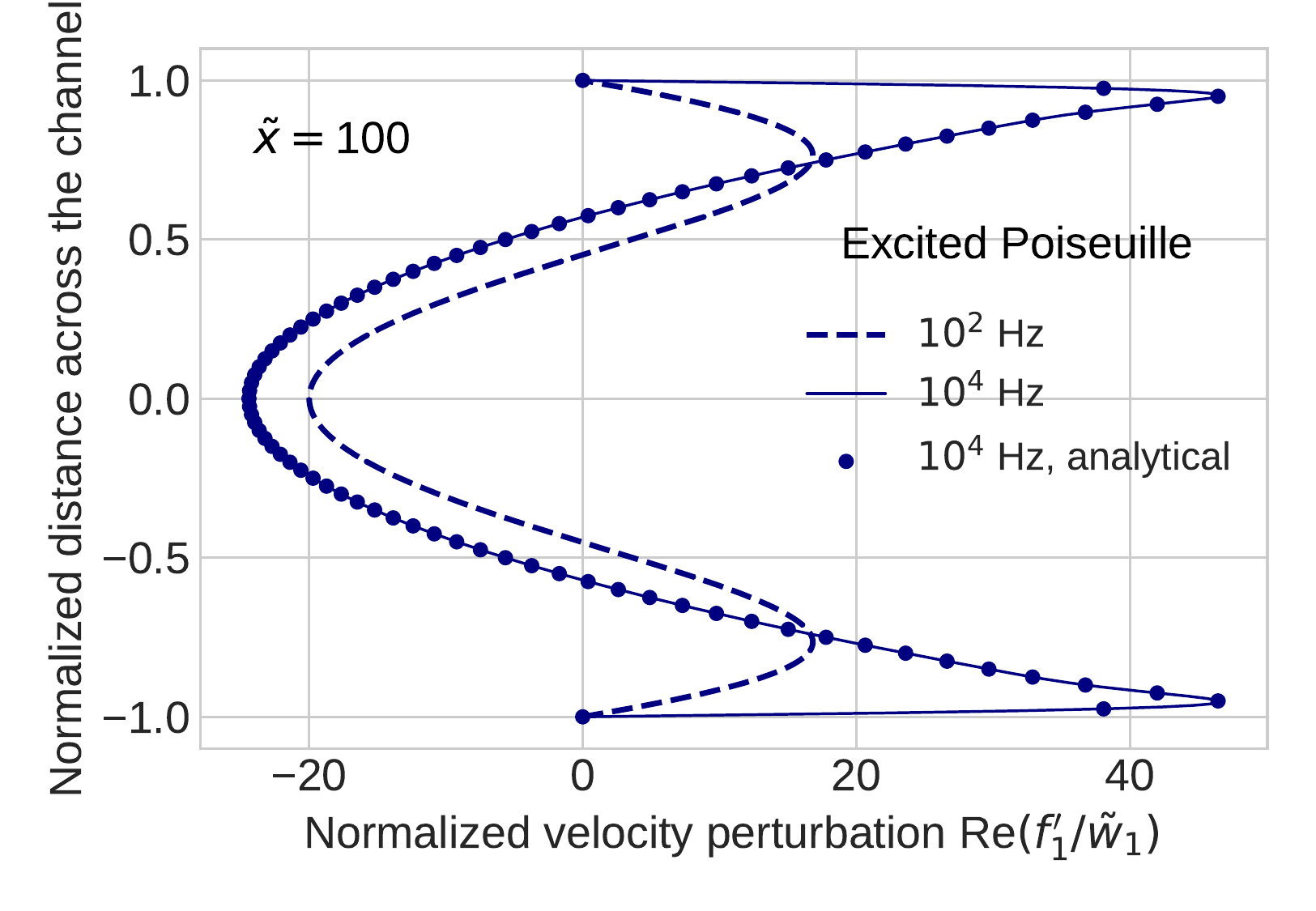}
        \caption{Transversal shape of longitudinal flow velocity
            oscillations amplitude for the transient Poiseuille problem,
            Eq.\eqref{eq:fPois}. Points show the analytical solution,
            Eq.\eqref{eq:f1psol}.
        }
        \label{fig:utransP}
    \end{center}
\end{figure}

The most interesting effect is dramatic growth of
the shoulders with the distance along the channel (Figure~\ref{fig:utrans}b).
Formally, the effect can be rationalized
considering the problem for the flow with zero
static injection velocity, however,
excited periodically at the permeable wall (excited Poiseuille flow).
An equation for the flow
velocity spectrum is obtained setting $\tw_0=0$ in Eq.\eqref{eq:fbert1}.
Further, at large frequencies, the term $\tw_1 \bigl(f_0'f_0' - f_0 f_0''\bigr)$
in Eq.\eqref{eq:fbert1} can be neglected, and we come to
\begin{equation}
    \ri\tom\bigl(\tx\tw_1 f_0' + f_1'\bigr) - \dfrac{f_1'''}{\Reyn} = k_1
    \label{eq:fPois}
\end{equation}
For Poiseuille flow, $f_0' = 3 (1 - \ty^2)/4$ and solution to
Eqs.\eqref{eq:fPois}, \eqref{eq:f1bc} leads to
\begin{multline}
    f_1' = \dfrac{\tx\tw_1}{4}
    \left(3\ty^2 - 1 - \dfrac{2\bigl(\phi\cosh(\phi\ty) - \sinh(\phi)\bigr)}
                             { \phi\cosh(\phi) - \sinh(\phi)}\right), \\
     \phi = \sqrt{\ri\tom\Reyn}
    \label{eq:f1psol}
\end{multline}
Eq.\eqref{eq:f1psol} shows that the longitudinal velocity perturbation
amplitude is proportional to the distance $\tx$.
The numerical transversal ($\ty$--) dependence of the perturbation amplitude for
the three frequencies is shown in Figure~\ref{fig:utransP}.
The curves are symmetric with respect to the mid--plane;
nonetheless, the trend with the frequency growth is the same as in Figure~\ref{fig:utrans}:
a shape with the two shoulders forms (Figure~\ref{fig:utransP}).
Note that the shoulders in Figure~\ref{fig:utransP} are twice
larger than those in Figure~\ref{fig:utrans}b, i.e., zero injection
velocity enhances the effect. Points in Figure~\ref{fig:utransP}
show the analytical solution, Eq.\eqref{eq:f1psol}.

At large frequencies,
the term with hyperbolic functions in Eq.\eqref{eq:f1psol}
has a sharp peak in a narrow domain near the walls, at $1 - |y| \lesssim \epsilon$.
The dimensionless width $\epsilon$
of this domain is on the order of $1/|\phi| = 1/\sqrt{\tom\Reyn}$,
which in the dimension form gives the distance between the shoulder
and the wall, Eq.\eqref{eq:last}.

% \subsection{Pressure gradient}

To understand the physics behind enhancement of velocity oscillations
amplitude along the channel, consider equation \eqref{eq:fx1} for the pressure gradient.
Setting $R=\tw\tx$, neglecting the term with $\tw'$ and
performing linearization and Fourier--transform, we come to
\begin{multline}
    - \pdr{\tp_1}{\tx} =
    \ri\tom\bigl((1 + \tx\tw_0) f_1' + \tx\tw_1 f_0'\bigr) \\
    + \tw_1\left(1 + 2\tx\tw_0\right)\left(f_0'f_0' -  f_0 f_0''\right) \\
    + (1 + \tx\tw_0) \tw_0\left(2 f_0'f_1' -  f_1 f_0'' - f_0 f_1''\right) \\
    - \dfrac{1}{\Reyn}\left((1 + \tx\tw_0) f_1''' + \tx\tw_1 f_0'''\right)
    \label{eq:dp1dx}
\end{multline}
where $\tp = \tp_0(\tx,\ty) + \tp_1(\tom,\tx,\ty)\exp(\ri\tom\tit)$ has been substituted.

\begin{figure}
    \begin{center}
        \includegraphics[scale=0.45]{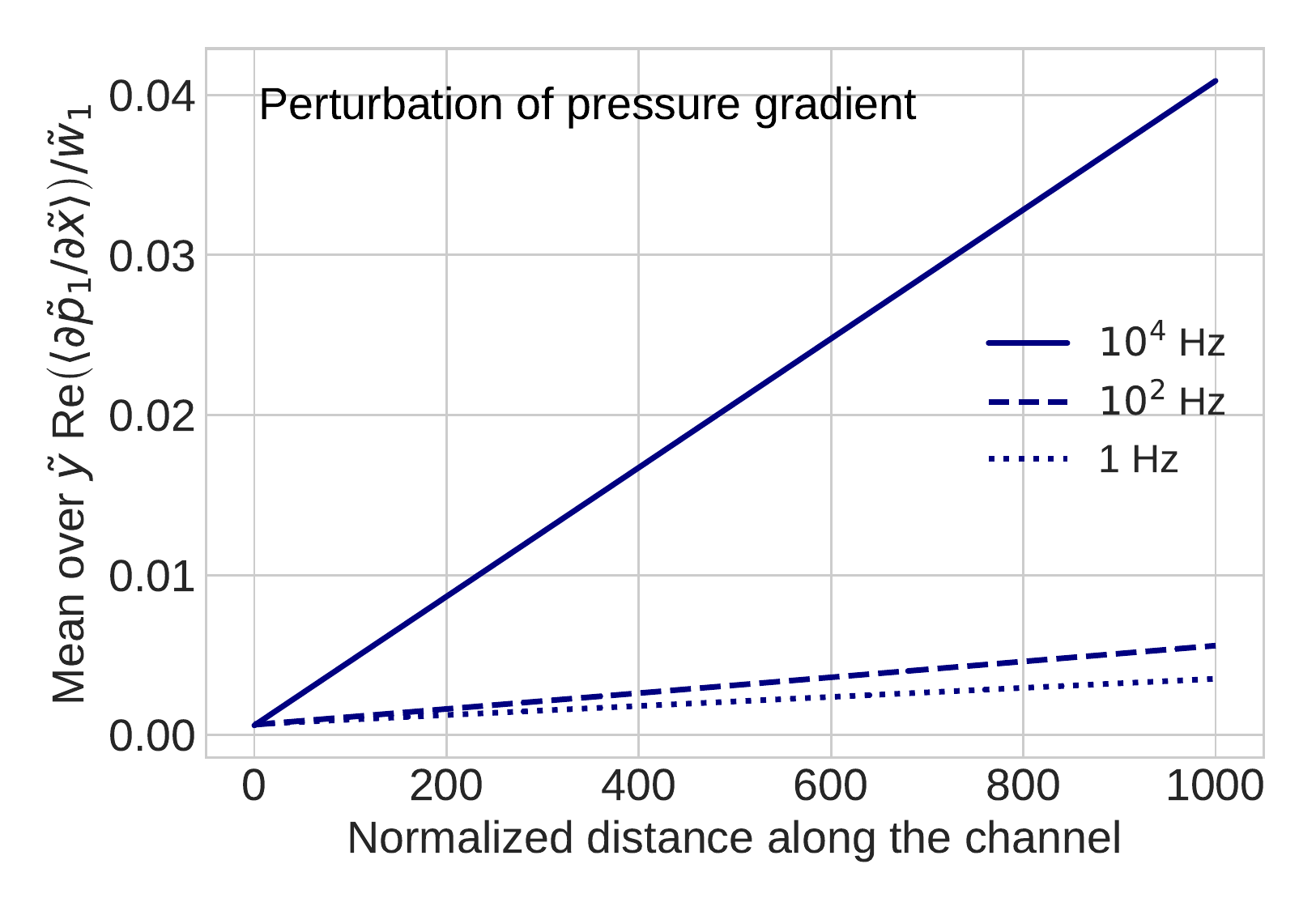}
        \caption{ The real component of pressure gradient perturbation
            amplitude along the channel. Perturbation amplitude
            of injection velocity is $10^{-3}$ m~s$^{-1}$.
        }
        \label{fig:dpdx}
    \end{center}
\end{figure}

With the parameters from Table~\ref{tab:parms}, the average over $\ty$--axis
real part of normalized pressure gradient perturbation
\begin{equation}
   -\dfrac{1}{\tw_1}\Braket{\pdr{\tp_1}{\tx}}
       \equiv - \dfrac{1}{2\tw_1}\int_{-1}^1 \pdr{\tp_1}{\tx}\, d\ty
\end{equation}
increases linearly
along the channel (Figure~\ref{fig:dpdx}). This explains the growth
of oscillation amplitude with $\tx$.
Indeed, the Sexl model \citep{Sexl_1930} for perturbed flow
in the channel with impermeable walls
results in the velocity oscillations amplitude proportional to the
amplitude of pressure gradient perturbation.
In the channel with mass injection,
a linearly increasing perturbation of pressure gradient forms, leading
to growing amplitude of velocity oscillations.

This result can be shown directly. From the first of Eqs.\eqref{eq:tutv}
we can write
\begin{multline}
    \tu_0 + \tu_1 = (1 + R_0 + R_1)(f'_0 + f'_1) \\
    \simeq (1 + R_0) f'_0 + (1 + R_0)f'_1 + R_1 f'_0
    \label{eq:tuex}
\end{multline}
Since $\tu_0 = (1 + R_0) f'_0$, from Eq.\eqref{eq:tuex} we find
\begin{equation}
    \tu_1 = (1 + R_0)f'_1 + R_1 f'_0
    \label{eq:tu1}
\end{equation}
Integrating the last equation over $\ty \in [-1,1]$ and taking into account
the boundary conditions \eqref{eq:fbc}, \eqref{eq:f1bc}, we find $\Braket{\tu_1} = R_1/2$.
In terms of perturbation amplitudes the last equation reads
\begin{equation}
    \Braket{\tu_1} = \dfrac{\tw_1\tx}{2},
\end{equation}
i.e., the amplitude of average longitudinal flow velocity oscillations is independent
of frequency and it increases linearly with $\tx$.

Finally, we note that at finite $\tw_0$,
the shoulders in Figure~\ref{fig:utrans} grow with $\tx$
as long as $\tx\tw_0$ is small, as Eq.\eqref{eq:fbert1} shows.
In long channels, for $\tx\tw_0 \gg 1$ the dependence of Eq.\eqref{eq:fbert1}
on $\tx$ vanishes and the shoulders amplitude saturates. This dependence, however,
retains in the general case of variable with $\tx$ injection velocity.

\section{Conclusions}

A transient model of laminar incompressible flow between
parallel walls, one of which is
permeable to mass injection is developed. Two--dimensional
transient Navier--Stokes equations and continuity equation are reduced
to a single one--dimensional transient equation for the longitudinal velocity.
Linearization and Fourier--transform lead
to equation for the small perturbation amplitude of this velocity.

The results  show that a small harmonic perturbation of injection
velocity at the permeable wall is converted to oscillations
of longitudinal flow velocity. At high frequencies,
the transversal profile of oscillations amplitude
exhibits two peaks which dramatically increase along the channel.
The effect is caused by linearly growing along the channel
perturbation amplitude of pressure gradient.
The peaks are located at the distance $\sqrt{\nu/\omega}$ from the walls, i.e.,
with the frequency growth each peak moves toward the nearest wall.

\section*{Conflict of interest statement}

The author declares no conflict of interest.

\newpage

% \bibliographystyle{unsrtnat}
% \bibliography{Cell_08,Cell_0304,Cell_13,Cell_19,Cell,Impedance,MyPapers}

\begin{thebibliography}{12}
    \providecommand{\natexlab}[1]{#1}
    \providecommand{\url}[1]{\texttt{#1}}
    \expandafter\ifx\csname urlstyle\endcsname\relax
    \providecommand{\doi}[1]{doi: #1}\else
    \providecommand{\doi}{doi: \begingroup \urlstyle{rm}\Url}\fi

    \bibitem[Chao et~al.(2018)Chao, Shuili, Yufei, Zhengyang, Wangzhen, and
    Liumo]{Chao_18}
    G.~Chao, Y.~Shuili, S.~Yufei, G.~Zhengyang, Y.~Wangzhen, and R.~Liumo.
    \newblock A review of ultrafiltration and forward osmosis: {A}pplication and
    modification.
    \newblock \emph{{IOP} Conference Series: Earth and Environmental Science},
    128:\penalty0 012150, Mar 2018.
    \newblock \doi{10.1088/1755-1315/128/1/012150}.

    \bibitem[Eikerling and Kulikovsky(2014)]{EK_book_14}
    M.~Eikerling and A.~A. Kulikovsky.
    \newblock \emph{Polymer Electrolyte Fuel Cells: Physical Principles of
        Materials and Operation}.
    \newblock CRC Press, London, 2014.

    \bibitem[Berman(1953)]{Berman_53}
    A.~S. Berman.
    \newblock Laminar flow in channels with porous walls.
    \newblock \emph{J.\ Appl.\ Phys.}, 24:\penalty0 1232--1235, 1953.
    \newblock \doi{10.1063/1.1721476}.

    \bibitem[Terrill and Thomas(1969)]{Terrill_69}
    R.~M. Terrill and P.~W. Thomas.
    \newblock On laminar flow through a uniformly porous pipe.
    \newblock \emph{Appl. Sci. Res.}, 21:\penalty0 37--67, 1969.
    \newblock \doi{10.1007/BF00411596}.

    \bibitem[Terrill(1983)]{Terrill_83}
    R.~M. Terrill.
    \newblock Laminar flow in a porous tube.
    \newblock \emph{J.\ Fluids Eng.}, 105:\penalty0 303--307, 1983.
    \newblock \doi{10.1115/1.3240992}.

    \bibitem[Kosinski et~al.(1970)Kosinski, Schmidt, and Lightfoot]{Kosinski_70}
    A.~A. Kosinski, F.~P. Schmidt, and E.~N. Lightfoot.
    \newblock Velocity profiles in porous-walled ducts.
    \newblock \emph{Ind.\ Eng.\ Chem.\ Fundam.}, 9:\penalty0 502--505, 1970.
    \newblock \doi{10.1021/i160035a033}.

    \bibitem[Granger et~al.(1989)Granger, Dodds, and Midoux]{Granger_89}
    J.~Granger, J.~Dodds, and N.~Midoux.
    \newblock Laminar flow in channels with porous walls.
    \newblock \emph{Chem.\ Eng.\ J.}, 42:\penalty0 193--204, 1989.
    \newblock \doi{10.1016/0300-9467(89)80087-5}.

    \bibitem[Karode(2001)]{Karode_01}
    S.~K. Karode.
    \newblock Laminar flow in channels with porous walls, revisited.
    \newblock \emph{J.\ Membr.\ Sci.}, 191:\penalty0 237--241, 2001.
    \newblock \doi{10.1016/S0376-7388(01)00546-4}.

    \bibitem[Richardson and Tyler(1929)]{Richardson_1929}
    E.~G. Richardson and E.~Tyler.
    \newblock The transverse velocity gradient near the mouths of pipes in which an
    alternating or continuous flow of air is established.
    \newblock \emph{Proc. Phys. Soc.}, 42:\penalty0 1--15, 1929.
    \newblock \doi{10.1088/0959-5309/42/1/302}.

    \bibitem[Sexl(1930)]{Sexl_1930}
    T~Sexl.
    \newblock {\"U}ber den von {E. G.} {R}ichardson entdeckten {A}nnulareffekt.
    \newblock \emph{Z. Phys.}, 61:\penalty0 349--362, 1930.
    \newblock \doi{10.1007/BF01340631}.

    \bibitem[Harris et~al.(1969)Harris, Peevt, and Wilkinson]{Harris_69}
    J.~Harris, G.~Peevt, and W.~L. Wilkinson.
    \newblock Velocity profiles in laminar oscillatory flow in tubes.
    \newblock \emph{J.\ Phys.\ E: Sci.\ Instrum.}, 2:\penalty0 913--916, 1969.
    \newblock \doi{10.1088/0022-3735/2/11/301}.

    \bibitem[Warburg(1899)]{Warburg_1899}
    E.~Warburg.
    \newblock {\"U}ber das {V}erhalten sogenannter unpolarisirbarer {E}lectroden
    gegen {W}echselstrom.
    \newblock \emph{Ann. Physik und Chemie}, 67:\penalty0 493--499, 1899.
    \newblock \doi{10.1002/andp.18993030302}.

\end{thebibliography}

\newpage

\end{document}